\documentclass[showpacs,preprintnumbers,amsmath,amssymb,twocolumn,superscriptaddress,prl]{revtex4}

\usepackage{times}
\usepackage{graphicx}
\usepackage{amsfonts}
\usepackage{amsmath, amsthm, amssymb}

\newcommand{\ve}[1]{\mathbf{#1}}

\newcommand{\vk}{\ve{k}} 

\newcommand{\e}[1]{\mathrm{e}^{#1}}

\def\i{\mathrm{i}} 

\newcommand{\ie}{\textit{i.e. }}
\newcommand{\eg}{\textit{e.g. }}
\newcommand{\etal}{\emph{et al.}}
                    
\begin{document}
\title[Dirac-fermions and conductance-oscillations in $(s,d)$-wave superconductor/normal graphene junctions]{Dirac-fermions and conductance-oscillations in $(s,d)$-wave superconductor/normal graphene junctions}
\author{J. Linder}
\affiliation{Department of Physics, Norwegian University of
Science and Technology, N-7491 Trondheim, Norway}
\author{A. Sudb{\o}}
\affiliation{Department of Physics, Norwegian University of
Science and Technology, N-7491 Trondheim, Norway}
\date{Received \today}
\begin{abstract}
We investigate quantum transport in a normal/superconductor graphene heterostructure, 
including the possibility of an anisotropic pairing potential in the superconducting region. We find 
that under certain circumstances, the conductance displays an undamped, oscillatory behaviour as a function of applied bias voltage.
Also, we investigate how the conductance spectra are affected by a $d$-wave pairing symmetry. These 
results combine unusual features of the electronic structure of graphene with the unconventional 
pairing symmetry found for instance in high-$T_c$ superconductors.   
\end{abstract}
\pacs{74.45.+c, 74.78.Na}
\maketitle
Graphene is a monoatomic layer of graphite with a honeycomb lattice structure \cite{novoselov}.
The electronic properties of graphene display several intriguing features, such as a six-point 
Fermi surface and  Dirac-like low-energy energy dispersion around the Fermi-points. 
Condensed matter systems with such `relativistic' electronic structure properties constitute 
fascinating examples of low-energy emergent symmetries (in this case Lorentz-invariance). 
Another example where precisely this occurs is in one-dimensional interacting fermion systems,
where phenomena like breakdown of Fermi-liquid theory and spin-charge separation take place.   
Graphene features certain similarities to, but also important differences from, the nodal
Dirac fermions emerging in the low-energy sector of the pseudogap phase of $d$-wave superconductors 
such as the high-$T_c$ cuprates. When Lorentz-invariance emerges in the low-energy sector of 
higher-dimensional condensed matter systems, it is bound to attract much interest from 
a fundamental physics point of view. 
\par
Various aspects of resonant tunneling phenomena in N/N and N/I/N graphene structures have recently been investigated \cite{katsnelson}.
Although superconductivity does not appear intrinsically in graphene, it may nonetheless be induced by 
means of the proximity effect 
\cite{heersche}. Motivated by this, the authors of Refs. \cite{beenakker, sengupta} considered quantum transport in N/S and N/I/S graphene junctions for the case where the pairing potential is isotropic, leading to $s$-wave superconductivity. However, the hexagonal symmetry of the graphene lattice also admits unconventional 
order parameters such as $p$-wave or $d$-wave. The possible 
pairing symmetries on a hexagonal lattice up to $f$-wave pairing ($l=3$) was given in Ref.  
\cite{mazinjohannes}. Interestingly, among the allowed order parameters, one finds the 
$d_{x^2-y^2}$-symmetry, which is believed to be the dominant pairing symmetry in high-$T_\text{c}$ 
superconductors. Consequently, it should be possible to induce superconductivity with
nodes in the gap in graphene by manufacturing heterostructures of graphene and unconventional 
superconductors. It is of interest to investigate how this would affect coherent quantum transport 
in junctions with normal and superconducting graphene. In particular, it is essential to study 
possible zero-energy states (ZES) at the interface of such a junction. Such states are known to give rise 
to zero-bias conductance peaks (ZBCPs) in metallic N/S junctions \cite{tanaka}, and will influence the conductance
spectra of N/I/S junctions.
\par  
In this Letter, we take into account the possibility of an anisotropic pairing potential induced in graphene, 
and study coherent quantum transport in both N/S and N/I/S junctions. In addition, we show that in the latter structure, novel conductance-oscillations as a function of bias voltage are present both for $s$-wave and 
$d$-wave symmetry of the superconducting condensate due to the presence of low-energy `relativistic'
nodal fermions on the N-side. The period of the oscillations decreases with increasing 
width $w$ of the insulating region, and persists even if the Fermi energy in I is strongly shifted. 
This contrasts sharply to metallic N/I/S junctions, where the presence of a potential barrier causes the 
transmittance of the junction to go to zero with increasing $w$. The feature of conductance-oscillations 
is thus unique to N/I/S junctions with low-energy Dirac-fermion excitations. Moreover,
we contrast the N/S or N/I/S conductance spectra for the cases where $s$-wave and $d_{x^2-y^2}$-wave 
superconductor constitutes the S-side. The former has no nodes in the gap and lacks Andreev bound 
states. The latter has line-nodes that always cross the Fermi surface in the gap, and thus features
in addition to Andreev bound states, also nodal relativistic low-energy Dirac fermions. 
The quantum transport properties in a heterostructure of two such widely disparate systems,
both featuring a particular intriguing emergent low-energy symmetry, is of considerable importance.  
\par
The Brillouin zone of graphene is hexagonal and the energy bands touch the Fermi level at the edges of this 
zone, amounting to six discrete points. 
Out of these only two are inequivalent, denoted $K$ and $K'$ and referred to as Dirac points. The energy 
dispersion in the Brillouin zone was calculated within a tight-binding model \cite{wallace}, 
revealing a conical structure of the conduction and valence bands close to the six Fermi points, giving 
rise to an essentially linear dispersion. 
Graphene N/S interfaces contain a new phenomenology compared to their metallic counterpart, namely the 
possibility of \textit{specular} Andreev-reflection (AR) \cite{beenakker}. In the process of normal AR, 
an incident electron from the N side is reflected as a hole which retraces the trajectory of the electron. 
In specular AR, the reflected hole follows the trajectory which a normally reflected electron would have. 
Depending on whether the graphene is doped or not, specular and normal AR will compete with each other, 
also depending on the position of the Fermi level with respect to the gap. 
\par
In order to treat the scattering 
processes at the interfaces of the N/I/S junction, we make use of the full Bogoliubov-de Gennes (BdG) 
equation for the 2D sheet of graphene in the $xy$-plane, assuming the clean limit. These equations read
\begin{align}\label{eq:Bdg1}
\begin{pmatrix}
\check{H} - E_\text{F}\check{1} & \Delta_\vk \check{1} \\
\Delta_\vk^\dag \check{1} & E_\text{F}\check{1} - \check{\mathcal{T}}\check{H}\check{\mathcal{T}}^{-1}\\
\end{pmatrix}
\Psi = E \Psi
\end{align}
where $E$ is the excitation energy, and $\Psi$ is the wave-function. We use $\check{...}$ for $4\times4$ matrices, $\hat{...}$ for $2\times2$ matrices, and 
boldface notation for three-dimensional row vectors. Assuming that the superconducting region is located 
at $x>0$ and neglecting the decay of the order parameter in the vicinity of the interface \cite{bruder}, 
we may write for the spin-singlet order parameter $\Delta_\vk = \Delta(\theta)\e{\i\vartheta}\Theta(x),$
where $\Theta(x)$ is the Heaviside step function, while $\vartheta$ is the phase corresponding the globally 
broken $U(1)$ symmetry in the superconductor. We consider the weak-coupling limit with the momentum $\vk$ 
fixed on the Fermi surface, such that $\Delta_\vk$ only has an angular dependence $\theta = \text{atan}(k_y/k_x)$. 
Note that in contrast to previous work, we allow for the possibility of unconventional superconductivity in 
the graphene layer since $\Delta_\vk$ now may be anisotropic. 
\par
Postulating a spin-singlet even parity order parameter, the condition $\Delta(\theta) = \Delta(\pi+\theta)$ 
must be fulfilled. The single-particle Hamiltonian is given by 
$\check{H} = \text{diag}(\hat{H}_+,  \hat{H}_-)$, $
\hat{H}_\pm = -\i v_\text{F} (\hat{\sigma}_x\partial_x \pm \hat{\sigma}_y\partial_y).$ Here, $v_\text{F}$ 
is the energy-independent Fermi velocity for graphene, while $\hat{\sigma}_i$ denotes the Pauli matrices. 
For later use, we also define the Pauli matrix vector 
$\hat{\boldsymbol{\sigma}} = (\hat{\sigma}_x, \hat{\sigma}_y, \hat{\sigma}_z)$. These Pauli matrices operate 
on the sublattice space of the honeycomb structure, corresponding to the A and B atoms, while the $\pm$ sign 
refers to the two so-called \textit{valleys} of $K$ and $K'$ in the Brillouin zone. The spin indices may be 
suppressed since the Hamiltonian is time-reversal invariant. In addition to the spin degeneracy, there is 
also a valley degeneracy, which effectively allows one to consider either the one of the $\hat{H}_\pm$ 
set. We choose $\hat{H}_+$, and consider an incident electron from the normal side of the junction $(x<0)$ with 
energy $E$. For positive excitation energies $E>0$, the eigenvectors and corresponding momentum of the 
particles read $
\psi^\text{e}_+ = [1, \e{\i\theta}, 0,0]^\text{T}\e{\i p^\text{e}\cos\theta x}$, $p^\text{e} = (E +  E_\text{F})/v_\text{F}$,
for a right-moving electron at angle of incidence $\theta$, while a left-moving electron is described by the substitution $\theta\to\pi-\theta$. If Andreev-reflection takes place, a left-moving hole with energy $(-E)$ and angle of reflection $\theta_\text{A}$ is generated with belonging wave-function $\psi^\text{h}_- = [0,0,1,\e{-\i\theta_\text{A}}]^\text{T}\e{-\i p^\text{h}\cos\theta_\text{A} x}$, $p^\text{h} = (E - E_\text{F})/v_\text{F}$,
where the superscript e (h) denotes an electron-like (hole-like) excitation. Since translational invariance 
in the $\hat{\mathbf{y}}$-direction holds, the corresponding component of momentum is conserved. This condition 
allows for determination of the Andreev-reflection angle $\theta_\text{A}$ through $p^\text{h}\sin\theta_\text{A} = p^\text{e}\sin\theta.$ One thus infers that there is no Andreev-reflection ($\theta_\text{A} = \pm\pi/2$), 
and consequently no subgap conductance, for angles of incidence above the critical angle $\theta_\text{c} = \text{asin}[|E-E_\text{F}|/(E+E_\text{F})].$ 
\par
On the superconducting side of the system ($x>0$), the possible wavefunctions for transmission of a right-moving quasiparticle with a given excitation energy $E>0$ reads $
\Psi^\text{e}_+ = [u(\theta^+), u(\theta^+)\e{\i\theta^+},v(\theta^+)\e{-\i\phi^+},v(\theta^+)\e{\i(\theta^+-\phi^+)}]^\text{T}$, 
$\times\e{\i q^\text{e}\cos\theta^+ x}$, $q^\text{e} = (E'_\text{F} + \sqrt{E^2-\Delta^2})/v_\text{F}$, and 
$\Psi^\text{h}_- = [v(\theta^-), v(\theta^-)\e{\i\theta^-},u(\theta^-)\e{-\i\phi^-},u(\theta^-)\e{\i(\theta^--\phi^-)}]^\text{T}$
$\times\e{\i q^\text{h}\cos\theta^- x}$, $q^\text{h} = (E'_\text{F} - \sqrt{E^2-\Delta^2})/v_\text{F}.$
The coherence factors are given by $u(\theta) = [\frac{1}{2}(1 + \sqrt{E^2-|\Delta(\theta)|^2}/E)]^{1/2}$, $v(\theta) = [\frac{1}{2}(1 - \sqrt{E^2-|\Delta(\theta)|^2}/E)]^{1/2}$. Above, we have defined $\theta^+ = \theta_\text{S}^\text{e}$, $\theta^- = \pi-\theta_\text{S}^\text{h}$, and $\e{\i\phi^\pm} = \e{\i\vartheta}\Delta(\theta^\pm)/|\Delta(\theta^\pm)|$.
The transmission angles $\theta^\text{(i)}_\text{S}$ for the electron-like (ELQ) and hole-like (HLQ) 
quasiparticles are given by $q^\text{(i)}\sin\theta^\text{(i)}_\text{S} = p^\text{e} \sin\theta$, i$=$e,h. 
Note that for subgap energies $E<\Delta$, there is a small imaginary contribution to the wavevector, which 
leads to exponential damping of the wavefunctions inside the superconductor. For clarity, we have 
omitted a common phase factor $\e{\i k_yy}$ which corresponds to the conserved momentum in the $\hat{\mathbf{y}}$-direction. A possible Fermi vector mismatch (FVM) between the normal and superconducting region is accounted for by allowing for $E'_\text{F} \neq E_\text{F}$. The case $E'_\text{F} \gg E_\text{F}$ corresponds to a heavily doped superconducting region, while $E'_\text{F} = E_\text{F}$ describes undoped graphene. Since we are using a mean-field approach to describe the superconducting 
part of the Hamiltonian, it is implicitly understood that phase-fluctuations of the order parameter must be small. This amounts to imposing the restriction \cite{kleinert} $\xi/\lambda' \gg 1$, or equivalently, 
$E_\text{F}' \gg \Delta$. 
\par
The conductance of the N/I/S junction is given by \cite{btk} $G(eV)=G_\text{N}\int^{\theta_\text{c}}_{-\theta_\text{c}} $$\text{d}\theta\cos\theta $$[1 -|r(eV,\theta)|^2  $$+ P|r_\text{A}(-eV,\theta)|^2]$,
where $r$ and $r_\text{A}$ are the reflection coefficients for normal and Andreev reflection, respectively, $P=|p_\text{h}|\cos\theta_\text{A}/(|p_\text{e}|\cos\theta)$, 
while $G_\text{N} = \int^{\pi/2}_{-\pi/2} \text{d}\theta \cos\theta[4\cos^2\theta/(4\cos^2\theta + Z^2)]$ is a renormalization constant corresponding to the N/N metallic conductance \cite{kashiwaya96}. 
In this case, 
we have zero intrinsic barrier such that $Z=0$. We will apply the usual approximation $|r_\text{A}(-eV,\theta)| \simeq |r_\text{A}(eV,\theta)|$, which holds for subgap energies. Although it is not valid for energies above the gap, 
this is of little consequence for the final result, since Andreev reflection is suppressed for $eV>\Delta$. The reflection and transmission coefficients constitute a unitary scattering matrix, a property that essentially expresses a conservation of probability. In deriving the conductance, we have ensured that the scattering coefficients have been normalized by the incoming current through the factor $P$.
In order to obtain these coefficients, we make use of the boundary conditions $\psi|_{x=0} = \tilde{\psi}_ \text{I}|_{x=0}, \;\tilde{\psi}_\text{I}|_{x=d} = \Psi_\text{S}|_{x=d}$,
where we have defined the wavefunction in the insulating region $\tilde{\psi}_\text{I} = \tilde{t}_1\tilde{\psi}^\text{e}_+ + \tilde{t}_2\tilde{\psi}^\text{e}_- + \tilde{t}_3\tilde{\psi}^\text{h}_+ + \tilde{t}_4\tilde{\psi}^\text{h}_-.$
The wavefunctions $\tilde{\psi}$ differ from $\psi$ in that the Fermi energy is greatly shifted by means 
of \eg an external potential, such that $E_\text{F} \to E_\text{F}-V_0$ where $V_0$ is the barrier 
(equivalent to the role of $Z$ in Ref.~\onlinecite{btk}). The coefficients $r$ and $r_\text{A}$ may now 
be obtained by using the boundary conditions, but we leave the explicit calculations and somewhat 
cumbersome analytical results for a forthcoming paper.
\par
Consider first a N/I/S graphene junction. In the 
thin-barrier limit defined as $d\to0$ and $V_0\to\infty$ with $s$-wave pairing, Ref. \cite{sengupta} 
reported a $\pi$-periodicity of the conductance with respect to the parameter $\chi = V_0d/v_\text{F}$. 
In the present study, we do not restrict ourselves to isotropic pairing, nor to the thin-barrier limit, 
and show that new physics emerges from the presence of a finite-width barrier. We measure the width $w$
of region I in units of $d/\lambda_\text{F}$ and the potential barrier $V_0$ in units of $E_\text{F}$. 
The linear dispersion approximation is valid up to $\simeq 1$ eV \cite{wallace}, and we will consider 
typical Fermi energies  in graphene of $E_\text{F} = 100$ meV in the undoped case and a gap 
$\Delta = 1$ meV \cite{novoselov}. In the doped case, we set $E_\text{F}'= 10E_\text{F}$, and we 
also fix $V_0 = 10E_\text{F}$ in order to operate within the regime of validity of the linear 
dispersion approximation. The undoped situation originally refers to the case where the Fermi level 
is located at the Dirac point, although real experimental graphene samples will have free carriers, such that $E_\text{F}$ is pushed upwards. The \textit{doped} case 
denotes a large FVM between the N and S region which may be induced by chemical 
doping or by a gate voltage.
\par
\begin{figure}[h!]
\centering
\resizebox{0.5\textwidth}{!}{
\includegraphics{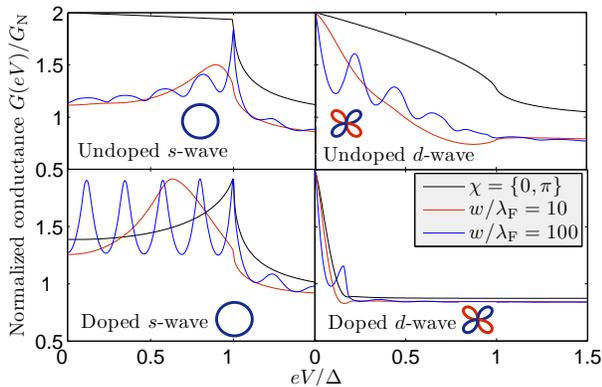}}
\caption{Tunneling conductance of N/I/S graphene junction for both $s$-wave and $d$-wave pairing in the undoped and doped case (see main text for parameter values). It is seen that for increasing $w$, a novel oscillatory behaviour of the conductance as a function of voltage is present in all cases.}
\label{fig:oscswave}
\end{figure}
Consider Fig. \ref{fig:oscswave} where we plot the normalized tunneling conductance in the two cases 
of $s$-wave and $d$-wave pairing, for both doped and undoped graphene. The most striking new feature 
compared to the thin-barrier limit is the strong oscillations in the conductance as a function of $eV$.
We also include the thin-barrier limit with $\chi=0$ and $\chi=\pi$ to illustrate the $\pi$-periodicity 
in this limit. For subgap energies, we regain the N/S conductance for undoped graphene when $\chi=0$, with nearly 
perfect Andreev reflection. To model the $d$-wave pairing, we have used the 
$d_{x^2-y^2}$ model $\Delta(\theta) = \Delta\cos(2\theta-2\alpha)$ with $\alpha=\pi/4$. The parameter 
$\alpha$ effectively models different orientations of the gap in $\vk$-space with regard to the 
interface, and $\alpha=\pi/4$ corresponds to perfect formation of ZES in N/S metallic 
junctions. For $\alpha=0$, the $d$-wave spectra are essentially identical to the $s$-wave case, since the condition for formation of ZES is not fulfilled in this case \cite{tanaka}. It is 
seen that in all cases shown in Fig. \ref{fig:oscswave} the conductance exhibits a novel oscillatory 
behavior as a function of applied bias voltage $eV$ as the width $w$ of the insulating region becomes 
much larger than the Fermi wavelength, \ie $w\gg \lambda_\text{F}$.
\par 
The oscillatory behavior of the conductance may be understood as follows. Non-relativistic free 
electrons with energy $E$ impinging upon a potential barrier $V_0$ are described by an expontentially 
decreasing non-oscillatory wavefunction $\e{\i kx}$ inside the barrier region if $E<V_0$, since the 
dispersion essentially  is $k\sim\sqrt{E-V_0}$. Relativistic free electrons, on the other hand, have a 
dispersion $k\sim (E-V_0)$, 
such that the corresponding wavefunctions do not decay inside the barrier region. Instead, the transmittance 
of the junction will display an oscillatory behavior as a function of 
the energy of incidence $E$. In general, 
a kinetic energy given by $\sim k^{\alpha}$ will lead to a complex momentum $k \sim (E-V_0)^{1/\alpha}$
inside the tunneling region, and hence damped oscillatory behavior of the wave function. Relativistic 
massless fermions are unique in the sense that only in this case ($\alpha = 1$) is the momentum purely 
real. Hence, the undamped oscillatory behavior at sub-gap energies appears as a direct manifestation
of the relativistic low-energy Dirac fermions in the problem. This observation is also linked to the so-called Klein paradox which occurs for electrons with such a relativistic dispersion relation, which has been theoretically studied in normal graphene \cite{katsnelson}.
\par
We next discuss why the illustrated conductance spectra are different for $s$-wave and $d$-wave symmetry, in 
addition to comparing the doped and undoped case. The doping level may be considered as an effective FVM, 
acting as a source of normal reflection in the scattering processes. This is why the subgap conductance at 
thin barrier limit is reduced in the doped case. Moving away from the thin barrier limit, it is seen 
that oscillations emerge in the conductance spectra. For $s$-wave pairing, the amplitude of the oscillations 
is larger in the doped case than in the undoped case, and the period of oscillations remains the same. This 
period depends on $w$, while the amplitude of the oscillations is governed by the wavevectors in the regions 
I and S. The maximum value of the  oscillations occurs when $2w$ equals an integer number of wavelengths,
corresponding to a constructive interference between the scattered waves. Physically, the amplitude-dependence 
of the oscillations on doping originates with the fact that doping effectively acts as an 
increase in barrier strength. By making $V_0$ larger, one introduces a stronger source of normal reflection. 
When the resonance condition for the oscillations is not met, the barrier  reflects the incoming particles 
more efficiently. This is also the reason why increasing $V_0$ directly and increasing $E_\text{F}'$ has 
the same effect on the spectra.
\par
We now turn to the difference between the $s$-wave and $d$-wave for the undoped case. It is seen that 
the conductance is reduced in the $d$-wave case compared to the $s$-wave case. One may understand the 
reduction in subgap conductance in the undoped case as a consequence of tunneling into the nodes of the 
gap, which is not present in the $s$-wave case. Hence, Andreev reflection which significantly contributes 
to the conductance, is reduced in the $d$-wave case compared to the $s$-wave case. Moreover, we see that 
a ZBCP is formed in the doped case, equivalent to a stronger barrier, and this is interpreted as the 
usual formation of ZES leading to a transmission at zero bias with a sharp drop for 
increasing voltage.
\begin{figure}[h!]
\centering
\resizebox{0.5\textwidth}{!}{
\includegraphics{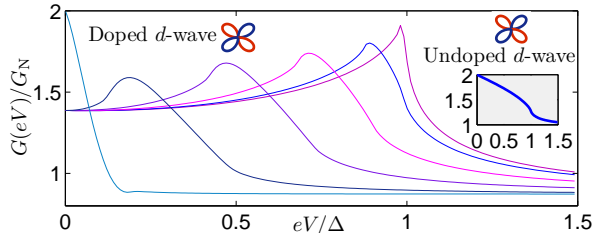}}
\caption{Conductance spectra for a doped N/S graphene junction in the $d$-wave case, varying the orientation angle $\alpha$. From right to left, the curves vary in the interval 
$\alpha\in[0,\pi/4]$ in steps of $\pi/20$. The inset shows the undoped case for $\alpha=\pi/4$.}
\label{fig:dwave}
\end{figure}
\par
Finally, we briefly investigate how the conductance spectra of a N/S graphene junction (without the insulating region) change when going from a $s$-wave to a $d$-wave order parameter in the superconducting part of the system. Consider Fig. \ref{fig:dwave} for the case of doped graphene, where we plot the conductance to see how it evolves upon a rotation of the gap. The behaviour is quite distinct from that encountered in a N/S metallic junction. From Fig. \ref{fig:dwave}, we see that the peak of the conductance shifts from $eV=\Delta$ to progressively lower values as $\alpha$ increases from $0$ to $\pi/4$, but only for $\alpha$ very close to $\pi/4$ is a ZBCP present. This is different from what is observed in metallic N/S junctions, where the formation of a ZBCP starts immediately as one moves away from $\alpha=0$ in the presence of a FVM, corresponding to the doped case here. In Fig. \ref{fig:dwave} the conductance spectra actually mimicks a lower value of the gap than what is the case, if one were to infer the gap magnitude from the position of the singularity in the spectra. This should be an easily observable feature in experiments, and provides a direct way of testing our theory. For undoped graphene, we found very little difference 
in the conductance spectra upon varying $\alpha$. The inset of Fig. \ref{fig:dwave} illustrates the undoped case for $\alpha=\pi/4$, where the deviation from perfect Andreev reflection for $eV<\Delta$ is due to tunneling into the nodes of the gap.
\par
In summary, we have studied coherent quantum transport in N/S and N/I/S graphene junctions, investigating 
also the role of $d$-wave pairing symmetry on the tunneling conductance. We report a new oscillatory behaviour of the conductance as 
a function of bias voltage for insulating regions that satisfy $d\gg \lambda$, which is present both for 
$s$- and $d$-wave pairing. In the latter case, we have also studied the conductance of an N/S junction and 
find very distinct behaviour from metallic N/S junctions: a rotation of $\alpha$ is accompanied by a 
progressive shift of the peak in the conductance. All of our predictions should be easily 
experimentally observable.
\par
\textit{Acknowledgments.} J. L. gratefully acknowledges D. Huertas-Hernando for very useful discussions in the initial stages of this work. This work was supported by the Norwegian Research Council Grants  No. 157798/432 and No. 158547/431 (NANOMAT), and Grant No. 167498/V30 (STORFORSK).

\end{document}